\documentclass[aps,prb,twocolumn,showpacs,preprintnumbers,amsmath,amssymb,superscriptaddress]{revtex4-1}

\usepackage{graphicx}
\usepackage{dcolumn}
\usepackage{bm}
\newcommand{\nc}{\newcommand}
\nc{\be}{\begin{equation}}
\nc{\ee}{\end{equation}}
\nc{\bea}{\begin{eqnarray}}
\nc{\eea}{\end{eqnarray}}
\nc{\bean}{\begin{eqnarray*}}
\nc{\eean}{\end{eqnarray*}}
\nc{\mb}{\mbox}
\nc{\rnc}{\renewcommand}
\nc{\vk}{\mb{\bf k}}
\nc{\vp}{\mb{\bf p}}
\nc{\vn}{\mb{\bf n}}
\nc{\vq}{\mb{\bf q}}
\nc{\rr}{\mb{\bf r}}
\nc{\vz}{\hat {\mb{\bf z}}}
\nc{\vj}{\mb{\boldmath$j$}}
\nc{\vg}{\mb{\boldmath$g$}}
\nc{\x}{\mb{\boldmath$x$}}
\nc{\A}{\mb{\boldmath$A$}}
\nc{\va}{\mb{\boldmath$a$}}
\nc{\vs}{\mb{\boldmath$\sigma$}}
\nc{\vpi}{\mb{\boldmath$\pi$}}
\nc{\nab}{\nabla}
\nc{\X}{\sf x}

\begin{document}

\title{Higgs Bosons in D-wave Superconductors}

\author{Yafis Barlas and C. M. Varma}
\affiliation{Department of Physics and Astronomy, University of California,
Riverside, CA 92521}

\maketitle
\begin{figure}
\begin{center}

\begin{tabular}{cc}
\includegraphics[width=1.7in,height=2.5in]{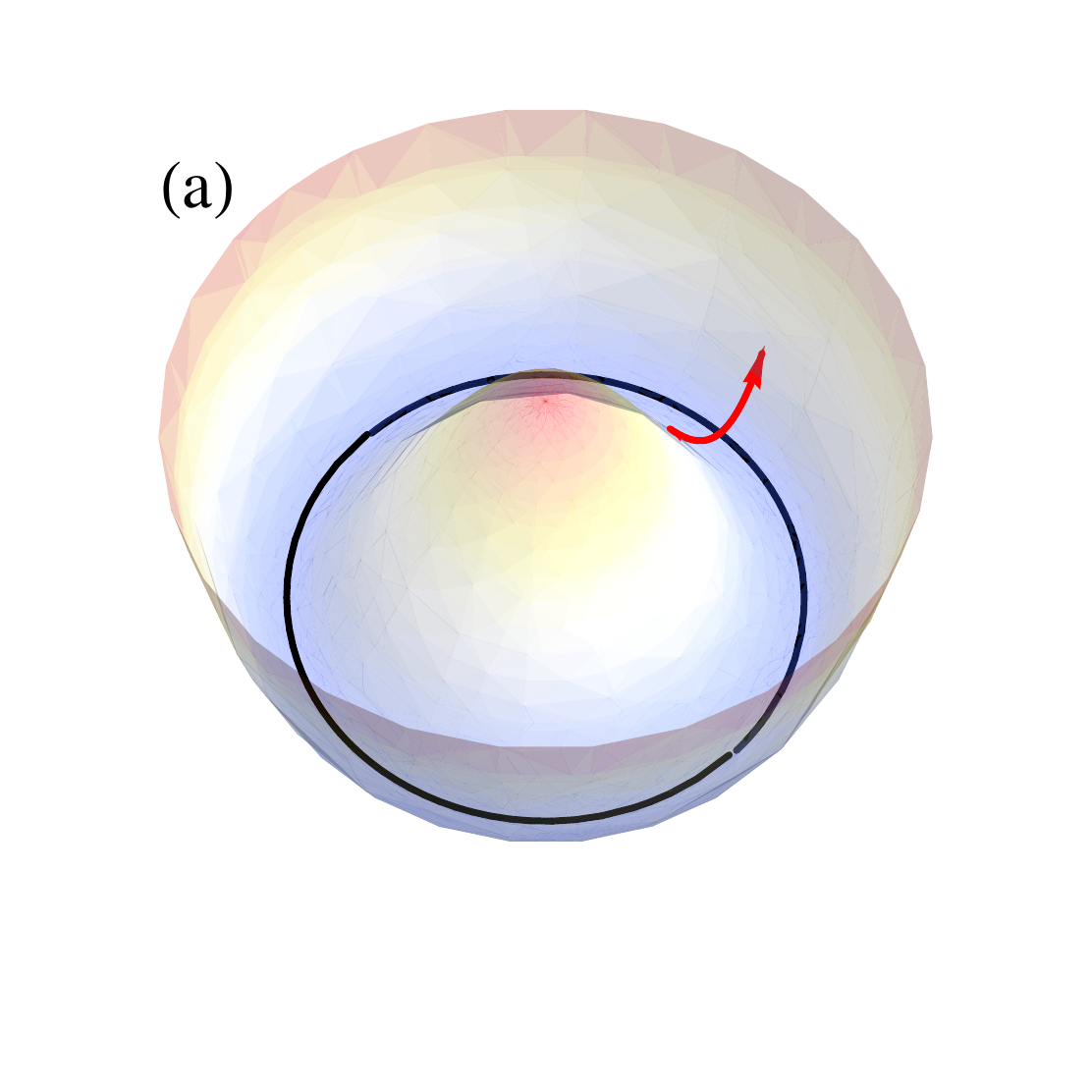} & 
\includegraphics[width=1.7in,height=2.5in]{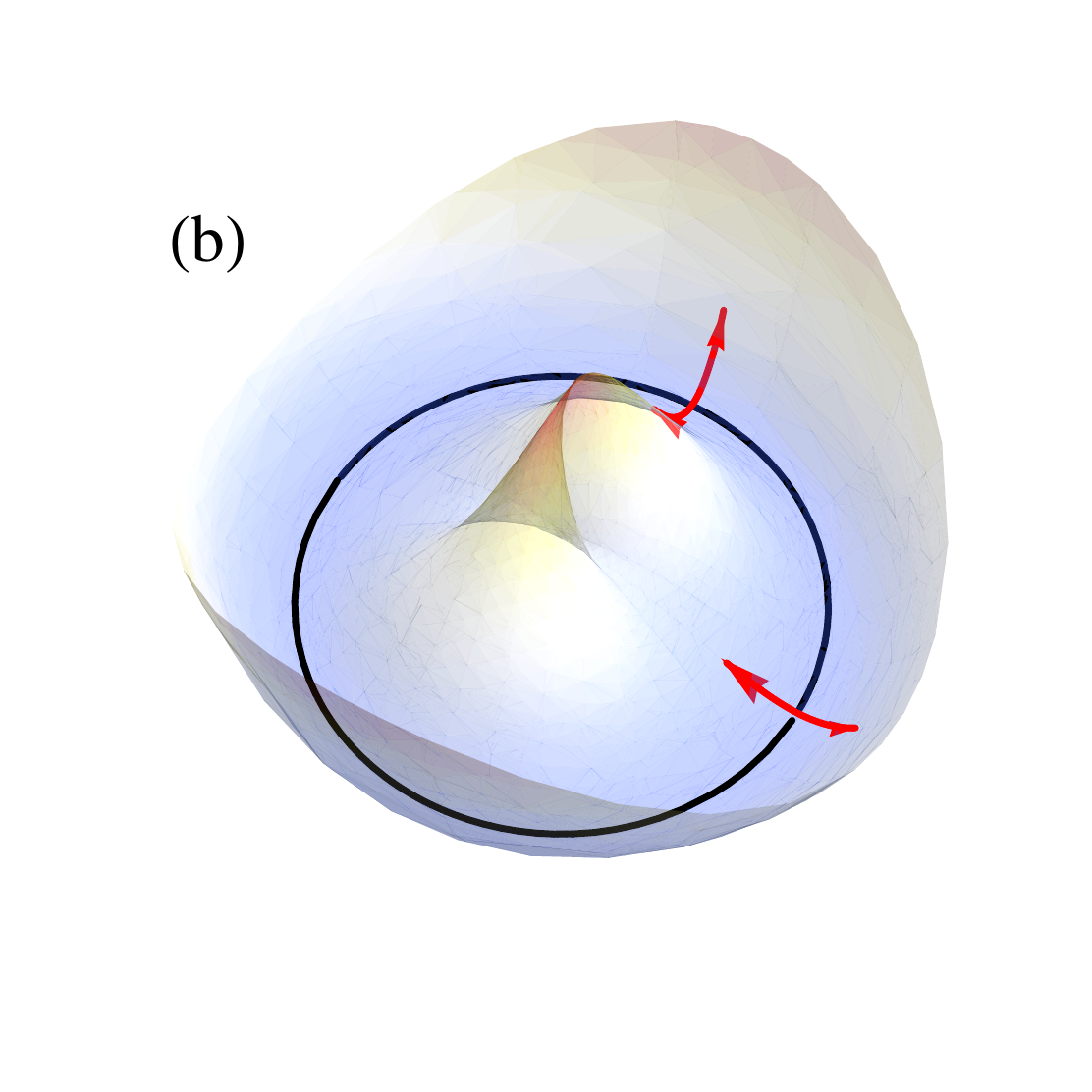} \\
\end{tabular}
\caption{Pictorial representation of the effective potential corresponding for a) s-wave Higgs mode and b) additional non s-wave Higgs modes.
The Nambu-Goldsone mode (black circle) oscillates in the azimuthal direction, whereas the amplitude (Higgs) mode (red curve) oscillates in the radial direction. In (a) the constant curvature of the effective potential results in an angle-independent finite mass gap. In (b) the effective potential plotted for the non-conventional Higgs modes $V(\rho_{i},\theta_{i}) = (a + b \sin^{2} \theta)(\rho_{i}^2 - \Delta_{i}^2)^2$ exhibits two-fold symmetry leading to a periodic angular dependent curvature. The non s-wave amplitude mode results in a more massive fluctuation at $\theta_{i} = \pi/2$ than $\theta_{i}=0$, in contrast with the s-wave Higgs mode. This leads to an angular dependent energy $\omega(\theta_{i})$ with the minimum and maximum occuring at $\theta_{i}=0,2\pi$ and $\theta_{i}=\pi/2,3 \pi/2$ respectively.} 
\label{figone}
\end{center}
\end{figure}
\begin{figure}
\begin{center}
\begin{tabular}{ccc}
\includegraphics[width=1.13in,height=1.13in]{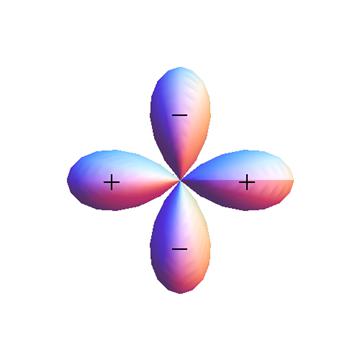} & 
\includegraphics[width=1.13in,height=1.13in]{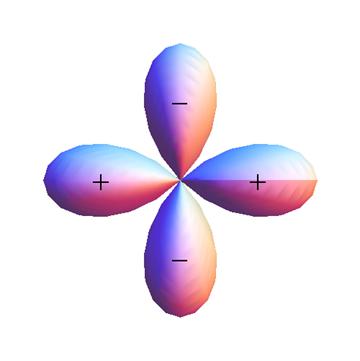} &
\includegraphics[width=1.13in,height=1.13in]{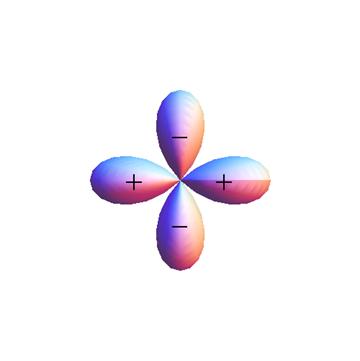}\\
\includegraphics[width=1.13in,height=1.13in]{unpert.jpg} & 
\includegraphics[width=1.13in,height=1.13in]{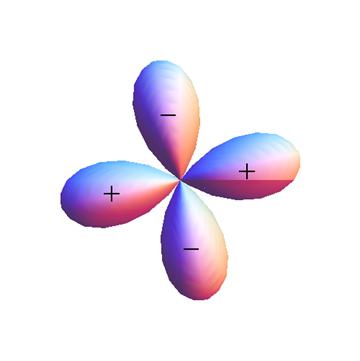} &
\includegraphics[width=1.13in,height=1.13in]{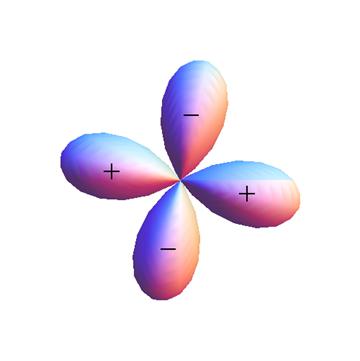}\\
\includegraphics[width=1.13in,height=1.13in]{unpert.jpg} & 
\includegraphics[width=1.13in,height=1.13in]{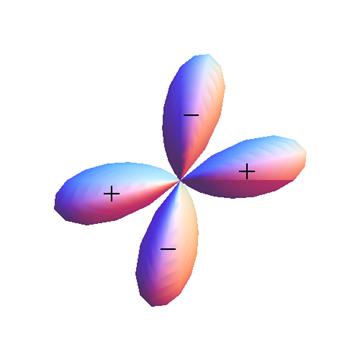} &
\includegraphics[width=1.13in,height=1.13in]{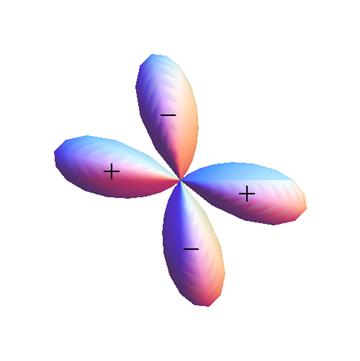}\\
\includegraphics[width=1.13in,height=1.13in]{unpert.jpg} & 
\includegraphics[width=1.13in,height=1.13in]{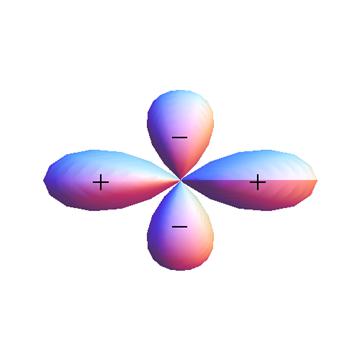} &
\includegraphics[width=1.13in,height=1.13in]{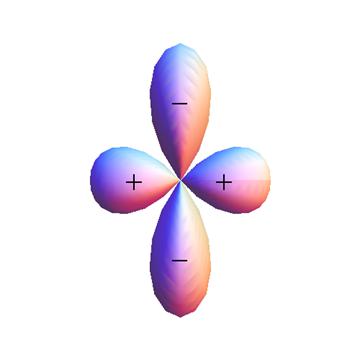}\\
\end{tabular}
\caption{Pictorial representation of the additional Higgs or amplitude modes of the d-wave superconducting order parameter predicted in this Letter. Each mode can be labeled by an irreducible representation of the point-group symmetry of the lattice in which the deformation of the order parameter occurs (see text for details). For the case of  $d_{k_{x}^2-k_{y}^2}$ order parameter depicted above these amplitude fluctuations are different admixtures of $d_{x^2-y^2}$-wave ("breathing mode"), $d_{xy}$-wave ("rotating mode"), $g_{xy(x^2-y^2)}$-wave ("clapping mode") and $s+g_{(x^2-y^2)^2}$-wave ("osculating mode") components to the ground state, labeled from top to bottom. }
\label{figtwo}
\end{center}
\end{figure}

{\bf The concept of "broken symmetry", that  the symmetry of the vacuum may be lower than the Hamiltonian of a quantum theory, plays an important role in modern physics. A manifestation of this phenomena is the Higgs boson in particle physics~\cite{hoddeson} whose long awaited discovery is imminent. An equivalent mode in superconductors is implicit in the early theories of their collective fluctuations~\cite{martin, Abrahams}. Spurred by some mysterious experimental results,~\cite{klein} the theory of the oscillation of the amplitude of superconductivity order parameter, which is the equivalent to the Higgs modes in s-wave superconductors~\cite{higgs} and its identification in the experiments, was explicitly provided~\cite{pbl-cmv}.  It was also shown that a necessary condition for this to occur~\cite{cmv} is the emergent Lorentz invariance in the superconducting state while the metallic state and the region just below $T_c$ is manifestly non-Lorentz invariant. Here we show that d-wave superconductors, such as the high temperature Cuprate superconductors, should have a rich assortment of Higgs bosons, each in a different irreducible representation of the point-group symmetries of the lattice.  We also show that these modes have a characteristic singular spectral structure which can be discovered in Raman scattering experiments.} 
\newline
\indent
The order parameter $\Psi$ in s-wave superfluids and superconductors is a complex number. The oscillation of the phase of $\Psi$ is the Bogolubov (Nambu-Goldstone) mode, which is massless at long wavelengths in a neutral superfluid. This phase mode may be understood as the azimuthal oscillation of a particle near the bottom of a Mexican hat potential, depicted in Fig. \ref{figone}a. In a charged superconductor, it moves to the frequency of the plasmon in a gauge invariant theory coupling phase modes to electromagnetism~\cite{Anderson}. While very interesting as the $W$-boson in particle physics, it tells nothing new about excitations of the superconducting state. In contrast, the amplitude mode, which oscillates in the radial direction does not couple to charge and has an excitation gap (mass gap) at long wavelengths equal to twice the superconducting gap $2 \Delta$. This is just where the continuum of particle-hole excitations begin, hence it is heavily damped and usually unobservable. Special situations which lower its energy are therefore required to detect this mode in s-wave superconductors.~\cite{pbl-cmv} Interesting related modes have also been discussed in superfluid $^3He$ \cite{wolfe}.   \newline
\indent
In this Letter we show that superconductors with lower symmetries support additional amplitude or Higgs modes labeled by the point group symmetry in which the deformation of the order parameter occurs. As expected, one of these modes is the conventional s-wave Higgs mode which appears at $2 \Delta$ and is likely to be overdamped.~\cite{pbl-cmv} The other amplitude modes, however in general, have lower energies than $2\Delta$, where the damping is smaller. They may therefore be more easily identified. As these additional modes correspond to deformation of the ordered state to different irreducible representations with variable relative phases $\theta_{i}$, the Higgs modes acquire a characteristic singular two-peak lineshape which is derived here. \\
\indent
\indent
Besides the $U(1)$ gauge symmetry, anisotropic superconductors are also invariant under a point group symmetry determined by the crystal lattice structure.~\cite{revmodsigrist}  For definiteness, we consider a two-dimensional unconventional d-wave superconductor with $d_{x^2-y^2}$ ordering on a square lattice, point group symmetry $D_{4}$.\cite{dresselhaus} The high temperature Cuprate superconductors belong to this category. The nature of the additional amplitude modes is sketched in Fig.~\ref{figtwo} and corresponds to excited states with admixtures of additional $d_{x^2-y^2}$, $d_{xy}$-wave, $g_{xy(x^2-y^2)}$-wave and $s+g_{(x^2-y^2)^2}$-wave components to the ground state, respectively.  \\
\indent
We represent the ground state and the oscillations about it by the order parameter, 
\be
\Psi ({\bf Q},{\bf k}) =  \Psi_0({\bf k}) + \delta \Psi({\bf Q},{\bf k}, \omega) e^{i\theta({\bf Q})}.
\ee
$\Psi_0({\bf k})$ is the uniform ground state which we assume to be in the $B_{1g}$, i.e $(k_x^2-k_y^2)$ symmetry with phase $\theta = 0$. $\delta \Psi({\bf Q},{\bf k}, \omega)$ are the amplitude of the deviations representing the collective modes with total center of mass momentum ${\bf Q}$ and internal momentum ${\bf k}$ with phase $\theta({\bf Q})$.  At long wavelengths, $ \delta \Psi$ may be written as a separable function of ${\bf Q}$ and ${\bf k}$. The ${\bf k}$ dependence is expressed in the four one dimensional even parity irreducible representations $(B_{1g},A_{1g},B_{2g},A_{2g})$ of the $D_{4}$ point group symmetry. For notational simplicity, we will represent $\delta \Psi (0, {\bf k}) $ as linear combinations of   $\phi_i({\bf k}) = |\phi_i({\bf k})| \exp{(i \theta_i)}; i =0,1,2,3$, respectively.
In the limit ${\bf Q} =0$, the field theory is given by the Lagrangian (see supplemental material),
\begin{eqnarray}
\label{lagrangian}
\mathcal{L} &=& \sum_{i=0}^{3} |\partial_{t} \phi_{i}|^2 + a_{i} |\phi_{i}|^2 - b_{i} |\phi_{i}|^4 \\ 
&-& \nonumber
 \sum_{i<j} \bigg( c_{ij} |\phi_{i}|^2 |\phi_{j}|^2 + \frac{d_{ij}}{2} (\phi^{\star}_{i} \phi_{j} - \phi^{\star}_{j} \phi_{i})^2 \bigg).
\end{eqnarray}
We include only second order time-derivatives; this is only valid well below the Ginzburg-Landau regime near $T_c$, where a first derivative in time representing dissipation dominates and Higgs mode cannot occur due to lack of Lorentz invariance~\cite{cmv}. We have introduced two distinct set of parameters $c_{ij}$ and $d_{ij}$ in (\ref{lagrangian}) so that the energy of the collective modes depends on the relative phase $\theta_i$ between the assumed ground state representation and the others. This is required by symmetry and introduces distinctive features in the spectra of the collective modes as we see below.\\
\indent
The equations of motion using (\ref{lagrangian}) give the energy of the collective modes at ${\bf Q} =0$ to be
\begin{equation}
\label{contenergies}
\omega_{i} (\theta_{i}) = \pm \sqrt{(c_{i0}+d_{i0} \sin^{2}(\theta_{i}))|\Psi_{0}|^2 + a_{i}},
\end{equation}
here $\theta_i$ is the relative phase of the $i \ne 0$ order parameters with respect to the ground state order parameter $|\Psi_{0}|^2=-a_{0}/2b_{0}$. The $(k_x^2-k_y^2)$ order parameter assumed for $\Psi_0$ implies  $a_{0} < 0$ for $ T < T_{0}^{c} = T^{c}$ (where $T^{c}$ is the critical temperature).  $a_{i}'s (i \neq 0)$ remain positive as $T$ approaches $T^{c}$ from below. $c_{ij} > d_{ij} >0$'s are expected because of the competition between different order parameters. $\omega_{0} = \sqrt{4b_{0}|\Psi_{0}|^2}$ corresponds to the simple s-wave Higgs mode of the $d$-wave superconductor and appears at 2$\Delta$.  The energies $\omega_{i}$ correspond to  fluctuations of the $d_{k_{x}^2-k_{y}^2}$ order parameter in which it deforms to other point group symmetries as depicted in Fig. \ref{figtwo}. \\
\indent
The mass $\omega_{i}$ of the modes can be estimated from general considerations and by comparison with s-wave Higgs mode. In order to compare the energies $\omega_{i}$ with $\omega_{0}$ one can gain insight by using a two-parameter Landau-Ginzburg energy functional in the parameter subspace $(\phi_{0},\phi_{i})$. The phase diagram in this subspace allows for three broken symmetry phases a) $ |\phi_{0}|^2 =-a_{0}/(2b_{0}), |\phi_{i}|^2 =0$,  for $ a_{0} <0, a_{i} >0$ ; b) $|\phi_{0}|^2 =0; |\phi_{i}|^2 =-a_{i}/(2b_{i})$ for $ a_{0}>0, a_{i} <0$; and a mixed phase c) $|\phi_{0}|^2 \neq 0, |\phi_{i}|^2 \neq 0$ which only appears for $a_{i} <0 $ and $a_{0} <0$. Since we assume that the broken symmetry superconducting state has $d_{k_{x}^2-k_{y}^2}$ order, we must require that $ |a_{0}| > |a_{i}| $ for $a_{i} <0 $ and $a_{0} <0$. In order to avoid a second order transition to the mixed phase we must satisfy 
\begin{equation}
\label{lowerupper}
c_{i0} < 2 \sqrt{b_{0} b_{i}}, \quad {\rm{and}} \quad c_{i0} > 2b_{0}\frac{|a_{i}|}{|a_{0}|}, 
\end{equation}
which establishes an upper and a lower bound on the energies $\omega_{i}$.\newline
\indent
In order to estimate the values for $b_{i}$ and $a_{i}$ we assume an attractive potential is dominant for all the irreducible representations, 
\begin{eqnarray}
\nonumber
V(\vec{k}-\vec{k}') &=& V_{1} + V_{0}\alpha_{0}(\hat{k})\alpha_{0}(\hat{k}') + V_{3}\alpha_{2}(\hat{k})\alpha_{2}(\hat{k}')\\
 &+& V_{4} \alpha_{0}(\hat{k})\alpha_{2}(\hat{k})\alpha_{0}(\hat{k}')\alpha_{2}(\hat{k}'),
\end{eqnarray}
where $\alpha_{0}(\hat{k}) = \hat{k}_{x}^2 -\hat{k}_{y}^2$ and $\alpha_{2}(\hat{k}) = \hat{k}_{x}\hat{k}_{y}$, with $V_{i} < 0$ for all values of  $i$ and $|V_{0}| \gg |V_{i}| (i \neq 0)$. This is a natural assumption for the $d_{xy}$ symmetry, since the difference from $d_{x^2-y^2}$ arises only due to the anisotropy in the density of states, and so also for the $d_{xy(x^2-y^2)}$ case. No such strong argument can be given for the s-wave case and so a repulsive potential is allowed for this case.~\cite{footnote} In the "weak coupling" limit $N_{F}V_{0} << 1$, where $N_{F}$ is the density of states evaluated at the Fermi energy, an estimate for the values of $a_{i}$ and $b_{i}$ gives 
\begin{equation}
a_{i} = a_{0}\frac{V_{i}}{V_{0}} \frac{ \log (T/T^c_{i})}{(T - T^c)}, \qquad  b_{i} = b_{0}\frac{V_{i}}{V_{0}},
\end{equation}
for $T \sim T^c$. Since $a_{i} >0$ for $T\sim T^{c}$ the energies $\omega_{i}$ starting initially at a non-zero value decrease in magnitude for temperatures below the transition temperature $T^{c}$ as $a_{i} \to 0$ for $T \to T^c_{i}$, whereas $ \omega_{0}$ increases in magnitude as the temperature is lowered from the transition temperature $T^c$. This can be seen from a combination of energy expression (\ref{contenergies}) and the upper-bound on the values of $ c_{i0} \sim d_{i0} $ which gives $ c_{i0} < 2b_{0} (V_{i}/V_{0})^{1/2} << 4b_{0}$. This indicates that there exist temperatures $ T^{\star}_{i} < T^{c} $ where $ \omega_{i} << \omega_{0}$, thus establishing an upper bound on the energies $\omega_{i}$. \\
\indent  
\begin{figure}
\begin{center}
\includegraphics[width=3.4in,height=3.6in]{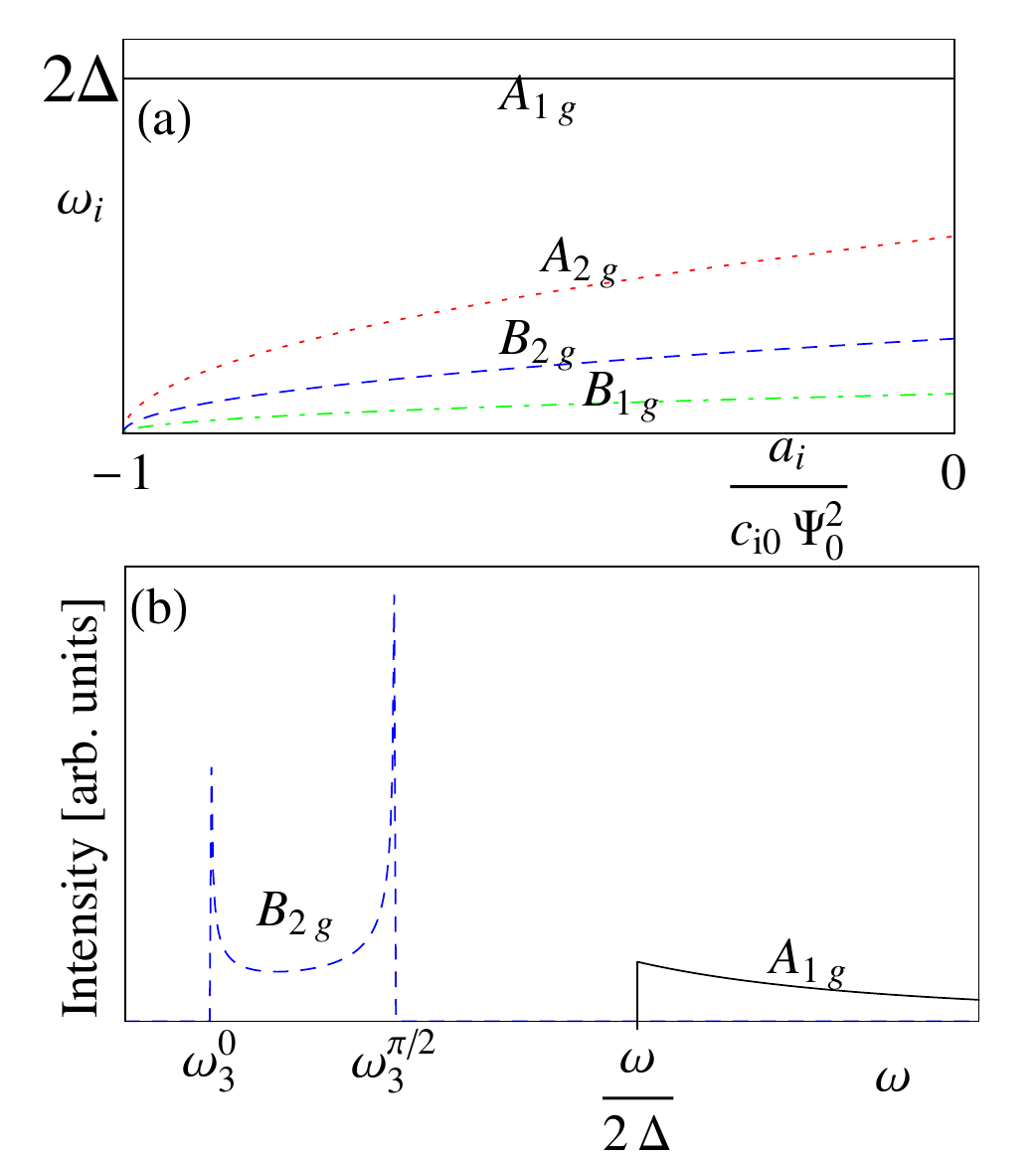}
\caption{(a) Schematic behavior of the energies for the additional Higgs modes at $\theta_{i}=0$ as a function of the coupling constants and superfluid density $a_{i}/c_{i0}|\Psi_{0}|^2$ at $T=0$. The modes are labeled by their Raman scattering geometry, the black (solid) line corresponds to the rotationally symmetric $A_{1g}$-Higgs mode, whereas the red(dotted), blue(dashed) and green(dot-dashed) lines correspond to the additional $A_{2g}$-,$B_{2g}$-,$B_{1g}$-Higgs modes(see text for details). We have assumed an attractive potential in all the irreducible representations.  (b) Schematic representation of the line shape associated to the "clapping" $B_{2g}$-Higgs mode (chosen arbitrarily) indicating the energy continuum and square-root singularities at the edges of the energy spectrum $\omega_{3}(0)$ and $\omega_{3} (\pi/2)$. The conventional 
$A_{1g}$-Higgs mode (which is likely to be overdamped) is shown as a reference.}
\label{figthree}
\end{center}
\end{figure}
\indent
The lower bound in Eq. \ref{lowerupper} follows from the condition that energies $\omega_{i}$ are always positive, so that no transition from the chosen ordered phase is allowed. The collective mode energies $\omega_{i}$ for the Higgs modes as a function of a combination of the phenomenological parameters and superfluid density are depicted schematically in Fig. \ref{figthree}a. An examination of Eq. (\ref{contenergies}) and subsequent considerations reveals that $\omega_i$  at $T=0$ for $(i \ne 0)$ is simply the difference of the ground state energy of the i-the symmetry from that of the realized $(i = 0)$ symmetry, as could have been guessed at the outset.\\
\indent
\indent
Since, at low energies, the quasi-particle density of states in a d-wave superconductor is proportional to the energy, the lower the energy of the modes in Fig.~\ref{figthree}a, the less they are damped. All the Higgs modes in d-wave superconductors, being oscillations of the amplitude of the superconducting condensate are neutral spin 0 modes. As such they do not couple to the usual external probes. In the case of s-wave superconductors \cite{pbl-cmv}, they could be discovered only through appearing in the self-energy of the superconducting state of phonons which promotes superconductivity, and steal intensity from them. Similarly, for cuprates, we expect that if the broad quantum-critical fluctuations, whose $q \to 0$ limit is  visible in Raman scattering, promote superconductivity \cite{ajishekhtervarma}, will partially give its weight to the Higgs modes. Elementary considerations indicate that $\omega_2$ or the breathing mode (which is likely to be over damped) occurs in the s-wave or $A_{1g}$ symmetry because for it $\delta\Psi$ also has $(x^2-y^2)$ symmetry,  the rotating mode occurs in the $A_{2g}$ symmetry because for it $\delta\Psi$ has $xy$ symmetry, the clapping mode occurs in the $B_{2g}$ symmetry because for it $\delta\Psi$ has $xy(x^2-y^2)$ symmetry and the osculating mode occurs in the $B_{1g}$ symmetry because for it $\delta\Psi$ has s-wave symmetry. The line shapes, which can be calculated from Eq.~\ref{contenergies}, exhibit a two peak structure with square root singularities at the edges of the energy spectrum as shown in Fig.~\ref{figthree}b (see supplementary material). The actual observation of the $A_{1g}$ mode may occur as a sharp peak below $2\Delta$ through coupling to the continuum. Indeed a mysterious intense mode in the $A_{1g}$ channel has already been detected \cite{sacuto}.\newline
\indent
In the supplemental material we deduce the Lagrangian for the gradient terms to derive the leading ${\bf Q}$ dependence of the energies. We find that there are interesting couplings between phase and amplitude modes,~\cite{chengwu} which are not influenced by Coulomb interactions, because of counterflow in the excited states of supercurrents in two different symmetries keep the system charge neutral. However, to quadratic order in ${\bf Q}$, the energy spectrum $\omega_{i}(\theta_{i})$ remains unchanged due to this coupling, only acquiring a quadratic dependence in the wavevector ${\bf Q}$ (see Eq. (10) in the supplementary material). Any effects of this coupling appear beyond quadratic order in ${\bf Q}$. \\
\indent
The low-energy physics of many condensed matter systems (lattice bosons near a Mott transition~\cite{coldatoms}, antiferromagnets~\cite{Haldane}, incommensurate charge-density wave and superconductors~\cite{Abrahams}), close to a quantum critical point, is captured by a Lorentz invariant critical theory~\cite{Sachdev}. One consequence of spontaneous breaking of a continuous symmetry in a Lorentz invariant theory is the appearance of amplitude fluctuation or "Higgs" modes. We show that when this symmetry is endowed with an additional discrete space symmetry, a rich assortment of Higgs modes should be present. The number of these Higgs modes should be equal to the number of irreducible representations of the discrete point group symmetry consistent with any internal symmetries (such as spin or valley) of the system.

Acknowledgements: We wish to thank Yuan Li and Alan Sacuto for discussion of Raman scattering results.


\begin{thebibliography}{99}
\bibitem{footnote}
A repulsive interaction in the s-wave channel will result in very large collective mode energy and will be completely overdamped.
\end{thebibliography}
\end{document}